\documentclass[twocolumn,aps,prl,groupedaddress,showpacs]{revtex4}
\usepackage{graphicx}
\usepackage{amsmath}
\usepackage{bm}

\begin{document}

\title{\mbox{}\\[10pt]
Production of the $X(3872)$ in $B$ Meson Decay\\
by the Coalescence of Charm Mesons}

\author{Eric Braaten and Masaoki Kusunoki}
\affiliation{
Physics Department, Ohio State University, 
Columbus, Ohio 43210, USA}
\author{Shmuel Nussinov}
\affiliation{
School of Physics and Astrononmy,
Tel Aviv University, Ramat Aviv, 
Tel Aviv 69978, Israel}

\date{\today}
\begin{abstract}
If the recently-discovered charmonium state $X(3872)$ is a 
loosely-bound S-wave molecule of the charm mesons 
$\bar D^0 D^{*0}$ or $\bar D^{*0} D^0$, it can be 
produced in $B$ meson decay by the coalescence of charm mesons.
If this coalescence mechanism dominates,
the ratio of the differential rate for $B^+ \to \bar D^0 D^{*0} K^+$
near the $\bar D^0 D^{*0}$ threshold and the rate for
$B^+ \to X K^+$ is a function of the $\bar D^0 D^{*0}$ 
invariant mass and hadron masses only.  
The identification of the $X(3872)$ as a
$\bar D^0 D^{*0}$/$\bar D^{*0} D^0$ molecule can be confirmed 
by observing an enhancement in the 
$\bar D^0 D^{*0}$ invariant mass distribution near the threshold. 
An estimate of the branching fraction 
for $B^+ \to X K^+$ is consistent with observations 
if $X$ has quantum numbers $J^{PC} = 1^{++}$ and if 
$J/\psi \; \pi^+ \pi^-$ is one of its major decay modes.
\end{abstract}



\maketitle


The recent unexpected discovery 
of a narrow charmonium resonance near 3.87 GeV 
challenges our understanding of heavy quarks and QCD.
This mysterious state $X(3872)$
was discovered by the Belle collaboration in electron-positron 
collisions through the $B$-meson decay $B^\pm  \to K^\pm X$
followed by the decay $X  \to J/\psi \; \pi^+ \pi^-$ \cite{Choi:2003ue}.
The discovery was confirmed by the CDF collaboration using
proton-antiproton collisions \cite{Acosta:2003zx}.
The $X$ is much narrower than all other charmonium states  
above the threshold for decay into a pair of charm mesons.
Its mass is also extremely close to the threshold for decay
into the charmed mesons $\bar D^0 D^{*0}$ or $\bar D^{*0} D^0$.

The proposed interpretations of the $X(3872)$
include a D-wave charmonium state with 
quantum numbers $J^{PC}= 2^{--}$ or $2^{-+}$,
an excited P-wave charmonium state 
with $J^{PC}= 1^{++}$ or $1^{+-}$,
a ``hybrid charmonium'' state in which a gluonic mode 
has been excited, and a $\bar D^0 D^{*0}$/$\bar D^{*0} D^0$  
molecule 
\cite{Tornqvist:2004qy,Close:2003sg,Pakvasa:2003ea,Voloshin:2003nt,%
Yuan:2003yz,Wong:2003xk,Braaten:2003he,Barnes:2003vb,Swanson:2003tb,%
Eichten:2004uh,Quigg:2004nv}. 
The possibility that charm mesons might form molecular states was 
considered some time ago \cite{Voloshin:ap,Nussinov:1976fg,Tornqvist:1991ks}.  
If the binding is due to pion exchange, the most favorable channels
are S-wave with quantum numbers $J^{PC} = 1^{++}$ or P-wave with $0^{-+}$
\cite{Tornqvist:2004qy}.
The proximity of the mass of $X$ to the $\bar D^0 D^{*0}$ threshold
indicates that it is extremely loosely bound.
If $X$ is an S-wave $\bar D^0 D^{*0}$/$\bar D^{*0} D^0$ molecule, 
the tiny binding energy introduces a new length scale,
the $\bar D^0 D^{*0}$ scattering length $a$,
that is much larger than other QCD length scales.
As a consequence, certain properties of the 
$X$/$\bar D^0 D^{*0}$/$\bar D^{*0} D^0$ system are determined by $a$ 
and are insensitive to the shorter distance scales of QCD.
This phenomenon is called {\it low-energy universality}. 

A challenge for any interpretation of the $X(3872)$
is to explain its production rate. 
This could be problematic for the identification of $X$ 
as an S-wave $\bar{D}^0D^{*0}$/$D^0\bar{D}^{*0}$ molecule, 
because it can readily dissociate due to its tiny binding energy.
One way to produce $X$ is to produce $\bar{D}^0$ 
and $D^{*0}$ with small enough relative momentum that they 
can coalesce into $X$. 
An example is the decay 
$\Upsilon(4S) \to X h h'$, where $h$ and $h'$ are light hadrons,
which can proceed through the coalescense into $X$ of charm mesons 
from the 2-body decays of a virtual $B$ and a virtual $\bar B$. 
Remarkably, low-energy universality determines
the decay rate for this process in terms of hadron masses 
and the width $\Gamma_B$ of the $B$ meson \cite{Braaten:2004rn}.  
Unfortunately, the rate is supressed by a factor of 
$(\Gamma_B/m_B)^2$ and is many orders of magnitude too small 
to be observed.

In this paper, we apply low-energy universality to the 
discovery mode $B^+ \to X K^+$ and to the process 
$B^+ \to \bar D^0 D^{*0} K^+$. 
We point out that the interpretation of $X$ as an S-wave 
$\bar D^0 D^{*0}$/$\bar D^{*0} D^0$ molecule can be confirmed 
by observing a peak in the $\bar D^0 D^{*0}$ invariant mass 
distribution near the $\bar D^0 D^{*0}$ threshold
in the decay $B^+ \to \bar D^0 D^{*0} K^+$.
We also estimate the branching fraction for $B^+ \to X K^+$.
The estimate is compatible with observations 
if $X$ has quantum numbers $J^{PC} = 1^{++}$ and if 
$J/\psi \; \pi^+ \pi^-$ is one of its major decay modes.

The mass of the $X$ has been measured to be 
$m_X = 3872.0 \pm 0.6 \pm 0.5$ MeV by Belle \cite{Choi:2003ue} and
$3871.4 \pm 0.7 \pm 0.4$ MeV by CDF \cite{Acosta:2003zx}.
It is extremely close to the $\bar D^0 D^{*0}$ threshold 
$3871.2 \pm 0.7$ MeV.
The binding energy is $E_b = -0.5 \pm 0.9$ MeV.  
If the state is bound, $E_b$ is positive, so
it is likely to be less than 0.4 MeV.
This is the smallest binding energy of any
S-wave two-hadron bound state.
The next smallest is the deuteron, 
a proton-neutron state with binding energy 2.2 MeV.
For two hadrons whose low-energy interactions are mediated by 
pion exchange, the natural scale for the binding energy of 
a molecule is $m_\pi^2/(2 \mu)$,
where $\mu$ is the reduced mass of the two hadrons.  
For a  $\bar D^0 D^{*0}$ molecule, this scale is 
about 10 MeV, so $E_b$ is at least an order
of magnitude smaller than the natural low-energy scale.

If the binding energy of $X$ is so small, low-energy universality 
implies that the $X$/$\bar D^0 D^{*0}$/$\bar D^{*0} D^0$ system
has properties that are determined by the 
$\bar D^0 D^{*0}$ scattering length $a$ and are 
insensitive to the shorter distance scales of QCD.  
The universal binding energy of the molecule is 
\begin{eqnarray}
E_b \equiv m_D+m_{D^*} - m_X \simeq \left( 2 \mu a^2 \right)^{-1},
\label{B2}
\end{eqnarray}
where $\mu = m_D m_{D^*}/(m_D+m_{D^*})$
is the reduced mass of the $\bar D^0$ and $D^{*0}$. 
The universal normalized momentum-space wavefunction 
at relative momentum $k \ll m_\pi$,
\begin{eqnarray}
\psi(k) \simeq (8\pi/a)^{1/2} (k^2+1/a^2)^{-1} ,
\label{psi}
\end{eqnarray}
was used by Voloshin 
to calculate the momentum distributions for the decays 
$X \to \bar D^0 D^0 \pi^0$ and $X \to \bar D^0 D^0 \gamma$ 
\cite{Voloshin:2003nt}.
The universal $\bar{D}^0 D^{*0}$ elastic scattering
amplitude at relative momentum $k_{\rm cm} \ll m_\pi$ is
\begin{equation}
{\cal A}[\bar{D}^0 D^{*0} \to \bar{D}^0 D^{*0}] \simeq 
\frac{8\pi m_D m_{D^*}}{\mu \left(-1/a-i k_{\rm cm} \right)},
\label{eq:scat}
\end{equation}
where $k_{\rm cm} \approx [2\mu(E - m_D - m_{D^*})]^{1/2}$
and $E$ is the total energy in the center-of-momentum frame.
The amplitude ${\cal A}[\bar D^{*0} D^0 \to \bar{D}^0 D^{*0}]$ 
for scattering to the CP conjugate state  differs by 
the charge conjugation $C = \pm$ of the channel 
with the large scattering length.
Another consequence of low-energy universality is that as 
the binding energy $E_b$ decreases, the probabilities for 
components of the wavefunction other than
$\bar D^0 D^{*0}$ and $\bar D^{*0} D^0$ decrease as $E_b^{1/2}$ 
\cite{Braaten:2003he}.
In the limit $E_b \to 0$, the state becomes 
$(|\bar D^{*0} D^0\rangle \pm |\bar D^0 D^{*0}\rangle )/\sqrt{2}$
if  $C = \pm$.
The rates for decays that do not correspond to the decay of a 
constituent $D^{*0}$ or $\bar D^{*0}$ also decrease as $E_b^{1/2}$.
This suppression may explain the surprisingly narrow width of the $X$.

\begin{figure}
\includegraphics[width=7cm]{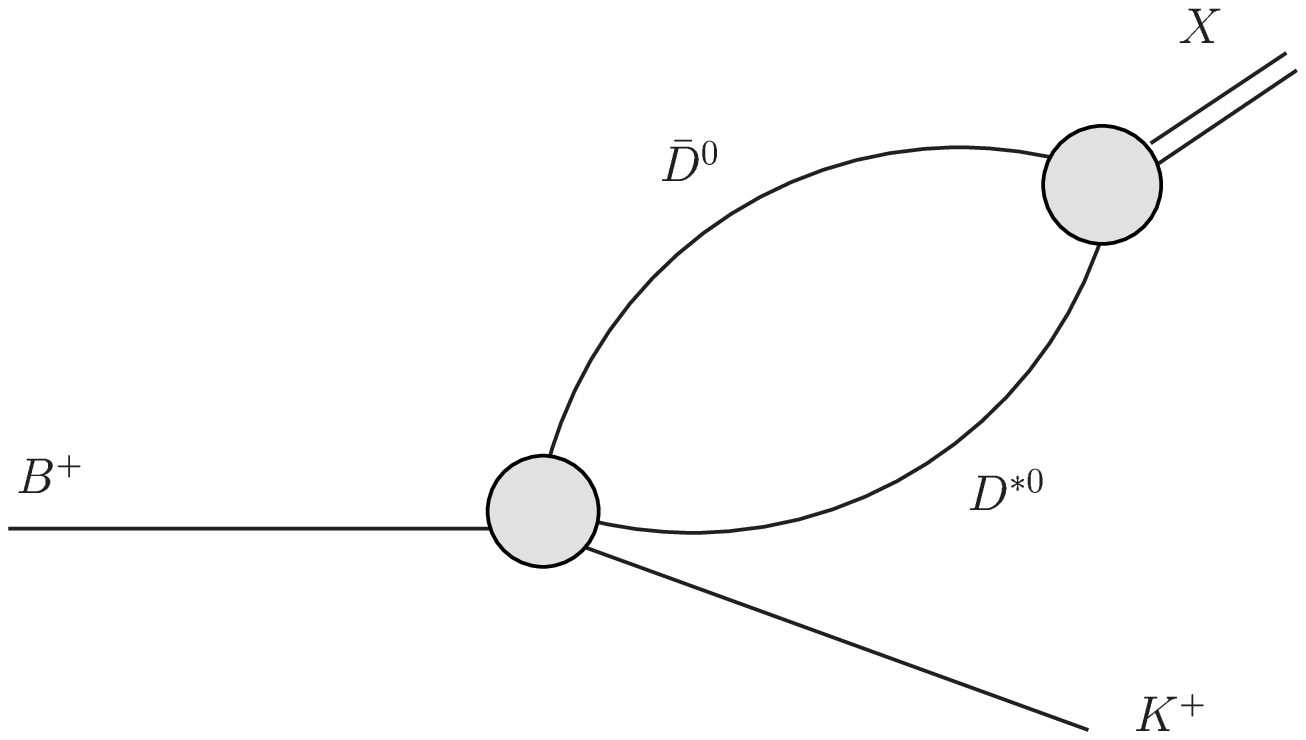}
\caption{Feynman diagram for 
$B^+ \rightarrow X K^+$ via the first pathway. 
\label{fig:B-XK}}
\end{figure}

The decay $B^+ \to X K^+$ proceeds through the weak decay
$\bar b \to \bar c c s$ at very short distances.
The subsequent formation of $X K^+$ is a QCD process
that involves momenta $k$ as low as $1/a$. The contributions
from $k \sim 1/a$ are constrained by low-energy universality, 
but those from $k \gtrsim m_\pi$ involve the full
complications of low-energy QCD. 
We analyze the decay $B^+ \to X K^+$ by separating 
short-distance effects involving  $k \gtrsim m_\pi$ 
from long-distance effects involving  $k \sim 1/a$.  
The decay can proceed via the short-distance
3-body decay $B^+ \to \bar{D}^0  D^{*0} K^+$ 
followed by the long-distance coalescence process 
$\bar{D}^0 D^{*0} \to X$. It can
also proceed through a second pathway consisting of 
$B^+ \to \bar D^{*0} D^0  K^+$ followed by $D^0  \bar{D}^{*0} \to X$. 
The amplitude for the first pathway can be expressed as
\begin{eqnarray}
&&{\cal A}_1[B^+\to X K^+] 
=  - i  {\sum} \int \! \! \frac{d^4\ell}{(2\pi)^4} \,  
{\cal A}[B^+\to  \bar D^0 D^{*0} K^+] 
\nonumber 
\\
&& \hspace{0.5cm} \times 
D(q+\ell,m_D) \, D(q_*-\ell,m_{D^*}) \, {\cal A}[ \bar D^0 D^{*0}\to X ],
\label{eq:amp1}
\end{eqnarray}
where $q = (m_D/m_X)Q$ and $q_* = (m_{D^*}/m_X)Q$
are 4-momenta that add up to the 4-momentum $Q$ of $X$ 
and $D(p,m) = (p^2 - m^2 + i \epsilon)^{-1}$.
The sum is over the spin states of the $D^{*0}$.
This amplitude can be represented by the Feynman diagram 
with meson lines shown in Fig.~\ref{fig:B-XK}.
We constrain the loop integral to the small-momentum region 
by imposing a cutoff $|\mbox{\boldmath  $\ell$}| < \Lambda$ 
in the rest frame of the virtual $D^0$ and $\bar D^{*0}$.
The natural scale for the cutoff is $\Lambda \sim m_\pi$.
The amplitude for  $\bar D^0 D^{*0}$ to coalesce into $X$ 
is determined by the $\bar D^0 D^{*0}$ scattering length $a$:
\begin{eqnarray}
&&{\cal A}[\bar D^0 D^{*0}\to X ] 
\nonumber
\\
&& \hspace{1cm}
= \left( 16\pi Z m_X m_D m_{D^*}/\mu^2 a \right)^{1/2} 
\epsilon_X^* \cdot \epsilon,
\label{eq:ampx}
\end{eqnarray}
where $\epsilon_X$ and $\epsilon$ are the polarization vectors 
of $X$ and $D^{*0}$ and $Z$ is the probability for the $X$
to be in a $\bar D^0 D^{*0}$/$\bar D^{*0} D^0$ state.  
At the $\bar D^0 D^{*0}$ threshold, the amplitude for 
$B^+ \to \bar D^0 D^{*0} K^+$ 
is constrained by Lorentz invariance to have the form
\begin{eqnarray}
{\cal A}[B^+ \to \bar D^0 D^{*0} K^+] 
= c_1 \, P \cdot \epsilon^*,
\label{eq:A-DD*K}
\end{eqnarray}
where $P$ is the 4-momentum of the $B$ meson and $c_1$ is a constant. 
The amplitude for 
$B^+ \to \bar D^{*0} D^0 K^+$ has the same form with $c_1$ replaced by 
a constant $c_2$.
In the $\bar D^0 D^{*0}$ rest frame, the integral over $\ell_0$ of the two 
propagators in (\ref{eq:amp1}) is proportional 
to the momentum-space wavefunction of $X$.
The subsequent integral over $\bm \ell$ is linear in the
ultraviolet cutoff $\Lambda$ for the low-momentum region:
\begin{eqnarray}
\int \!\! \frac{d^4\ell}{(2\pi)^4}\,  
D(q+\ell,m_D) \, D(q'-\ell,m_{D^*}) =
{i \mu \Lambda \over 4 \pi^2 m_D m_{D^*}} .
\end{eqnarray}
The total amplitude from the two pathways is 
\begin{eqnarray}
{\cal A}[B^+\to X K^+] = 
- \left( Z m_X/\pi^3 m_D m_{D^*} a \right)^{1/2}
\nonumber
\\
\times (c_1 \pm c_2) \Lambda \, P \cdot \epsilon_X^*.
\label{eq:ampsum}
\end{eqnarray}
The sign $\pm$ corresponds to the charge conjugation
$C = \pm$ of $X$.
Heavy-quark spin symmetry implies $c_1 = c_2$
up to corrections suppresssed by a factor $\Lambda_{\rm QCD}/m_B$.
The interference is constructive if $C = +$
and destructive if  $C = -$.
 The dependence of the loop amplitude (\ref{eq:ampsum}) on $\Lambda$ 
is cancelled by a tree diagram with a $B-XK$ contact interaction 
whose coefficient therefore depends linearly on $\Lambda$.  
If the $X$ is predominantly a $\bar D D^*$ molecule,
there must be some value $\Lambda_1$ of the ultraviolet cutoff 
for which the loop amplitude dominates over the tree amplitude. 
Squaring the amplitude, summing over spins, and integrating 
over phase space, the final result for the decay rate is
\begin{eqnarray}
\Gamma[B^+ \to X K^+] =
{Z \lambda^{3/2}(m_B,m_X,m_K)  \over 64 \pi^4 m_B^3 m_X^2 \mu a} 
	|c_1 \pm c_2|^2 \Lambda_1^2 ,
\label{eq:B-XK}
\end{eqnarray}
where $\lambda(x,y,z) = x^4 + y^4 + z^4 - 2 (x^2y^2+y^2z^2+z^2x^2)$. 
Due to the factor $1/a$, the decay rate scales like $E_b^{1/2}$
as $E_b \to 0$. 

If another hadronic state $H$ is
close enough to the $\bar D^0 D^{*0}$ threshold that $X$ 
has a nonnegligible probability $Z_H$ of being in the state $H$,
the decay can also proceed through a short-distance 2-body decay 
$B^+ \to H K^+$.  In this case, there is an additional 
term ${\cal A}[B^+ \to H K^+] Z_H^{1/2}$ in (\ref{eq:ampsum}).
Its contribution to the decay rate also scales 
like $E_b^{1/2}$ as $E_b \to 0$, 
because $Z_H$ scales like $E_b^{1/2}$ \cite{Braaten:2003he}.
If $C = +$, one possibility for such a state 
is the excited P-wave charmonium state $\chi_{c1}(2P)$.
Recent coupled-channel calculations of the charmonium spectrum
suggest that $\chi_{c1}(2P)$ is likely to be well above
the $\bar D^0 D^{*0}$ threshold \cite{Eichten:2004uh}.
We will henceforth assume that $\bar D^0 D^{*0}$/$\bar D^{*0} D^0$
is the only important component of the wavefunction 
and set $Z \approx 1$.

\begin{figure}
\includegraphics[width=7cm]{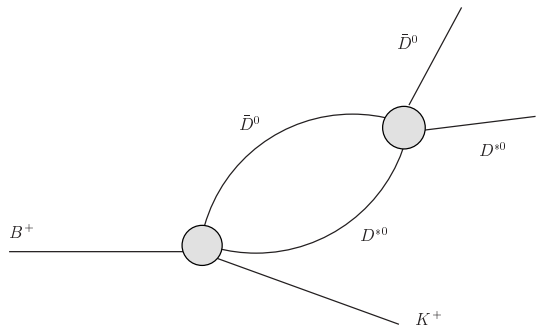}
\caption{Feynman diagram for 
$B^+ \rightarrow \bar D D^{*0} K^+$ via the first pathway. 
\label{fig:BDD*K}}
\end{figure}

\begin{figure}
\includegraphics[width=7cm,angle=270]{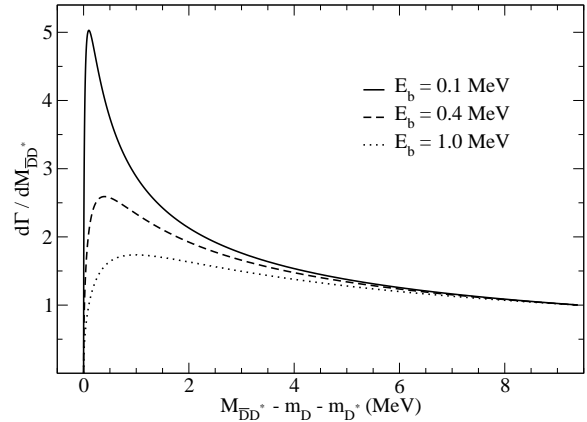}
\caption{The $\bar D^0 D^{*0}$ invariant mass distribution for 
$B^+ \rightarrow \bar D^0 D^{*0} K^+$ for three different values 
of the binding energy of $X$.
The distributions are normalized to 1 at $k_{\rm cm} = m_\pi$. 
\label{fig:MDD*}}
\end{figure}

We can calculate the differential decay rate 
for $B^+ \to \bar D^0 D^{*0} K^+$ in the same way.
There are again two pathways:
the short-distance decay $B^+ \to \bar{D}^0  D^{*0} K^+$ 
followed by the long-distance scattering
$\bar{D}^0 D^{*0} \to \bar{D}^0 D^{*0}$
and $B^+ \to \bar D^{*0} D^0 K^+$ 
followed by $\bar D^{*0} D^0 \to \bar{D}^0 D^{*0}$. 
The amplitude for the first pathway can be represented 
by the Feynman diagram with meson lines shown in Fig.~\ref{fig:BDD*K}.
The calculation of the amplitude is similar to that for 
$B^+ \to X K^+$ except that 
it involves the scattering amplitude (\ref{eq:scat})
instead of the coalescence amplitude (\ref{eq:ampx}).
In the loop amplitude for $B^+ \to \bar D^0 D^{*0} K^+$,
we keep only the term (\ref{eq:A-DD*K}) that is nonzero at the 
$\bar D^0 D^{*0}$ threshold.  
There must be some value $\Lambda_2$ of the ultraviolet cutoff 
for which the loop amplitude dominates over the tree amplitude. 
The factor $c_1 \pm c_2$ cancels in the ratio between the
amplitudes for $B^+ \to \bar D^0 D^{*0} K^+$ and $B^+ \to X K^+$. 
Our final expression for the differential decay rate is
\begin{eqnarray}
&& \frac{d\Gamma}{d M_{\bar D D^*}}[B^+\to \bar D^0 D^{*0} K^+]
\nonumber
\\
&& \hspace{1cm}
= \Gamma[B^+\to X K^+] \; {\Lambda_2^2 \over \Lambda_1^2} \;
{\mu a^3 k_{\rm cm} \over \pi (1 + a^2 k_{\rm cm}^2)} ,
\label{eq:B-DD*K}
\end{eqnarray}
where  $M_{\bar D D^*}$ is the $\bar D^0 D^{*0}$ invariant mass
and $k_{\rm cm}$ is the relative momentum
in the $\bar D^0 D^{*0}$ rest frame:
\begin{eqnarray}
k_{\rm cm} = 
\lambda^{1/2}(M_{\bar D D^*}, m_D, m_{D^*})/(2 M_{\bar D D^*}).
\end{eqnarray}
In (\ref{eq:B-DD*K}), we have neglected 
terms suppressed by $k_{\rm cm}^2/m_D^2$.
The invariant mass distribution is illustrated in Fig.~\ref{fig:MDD*}
for several values of the binding energy $E_b$.
The distributions are normalized to 1 at $k_{\rm cm} = m_\pi$.
As the binding energy is tuned toward 0, 
the peak value scales like $E_b^{-1/2}$ and the position of the peak 
in $M_{\bar D D^*} - (m_D + m_{D^*})$ scales like $E_b$.
The observation of such an enhancement near the $\bar D^0 D^{*0}$
threshold would confirm the interpretation of $X$ as a 
$\bar D^0 D^{*0}$/$\bar D^{*0} D^0$ molecule.

The Babar collaboration has recently measured the branching fractions 
for $B^+$ to decay into $\bar{D}^0 D^0 K^+$,
$\bar{D}^0 D^{*0}K^+$, $\bar D^{*0} D^0 K^+$, and $\bar D^{*0} D^{*0} K^+$
to be $(0.19 \pm 0.03)\%$, $(0.47 \pm 0.07)\%$, $(0.18 \pm 0.07)\%$,
and $(0.53 \pm 0.11)\%$, respectively \cite{Aubert:2003jq}.
We use these measurements to estimate 
the branching fraction for $B^+ \rightarrow X K^+$.
We make the simplifying assumption that the decay 
amplitude factors into currents $\bar c \gamma^\mu (1-\gamma_5) b$ and 
$\bar s \gamma^\mu (1-\gamma_5) c$.  Heavy quark symmetry can then 
be used to express the 3-body double-charm decay amplitudes 
in terms of two functions $G_1(q^2)$ and $G_2(q^2)$,
where $q^2$ is the invariant mass of the hadrons produced by the 
$\bar s \gamma^\mu (1-\gamma_5) c$ current \cite{Manohar-Wise}.   
For example, the 
amplitudes for decays into $\bar{D}^0 D^{*0} K^+$ 
and $\bar D^{*0} D^0 K^+$ are
\begin{eqnarray}
&& {\cal A}[B^+ \to \bar D^0 D^{*0} K^+] = 
-i G_1 \epsilon^* \cdot (V + v)
\nonumber
\\
&&  \hspace{0.5cm}
-i( G_2/m_B) \epsilon^*_\nu 
\left[ v_* \cdot k (V + v)^\nu 
	- v_* \cdot (V + v) k^\nu \right.
\nonumber
\\
&& \hspace{2.5cm} \left.
- i \epsilon^{\nu \mu \alpha \beta} (V + v)_\mu v_{* \alpha} k_\beta \right] ,
\label{A-Dbar*D}
\\
&& {\cal A}[B^+ \to \bar D^{*0} D^0 K^+] = 
i (G_1 v_\mu + G_2 k_\mu/m_B) \epsilon^*_\nu 
\nonumber
\\
&&  \hspace{0.5cm}
\times 
\left[ (1+ v_* \cdot V) g^{\mu \nu} 
	- v_*^\mu V^\nu
	-   i \epsilon^{\mu\nu \alpha \beta} v_{* \alpha} V_\beta \right] ,
\label{A-DbarD*}
\end{eqnarray}
where $k$ is the 4-momentum of the $K^+$ and $V$, $v_*$, and $v$ 
are the 4-velocities of the $B^+$, $\bar D^{*0}$ or $D^{*0}$, 
and $D^0$ or $\bar D^0$, respectively.
As a further simplification, we approximate $G_1$ and $G_2$
by constants.  The resulting expressions for the 
3-body double-charm decay rates are
\begin{eqnarray}
&& \Gamma[B^+ \to \bar D^0 D^0 K^+] = 10^{-3} {\rm MeV} 
\nonumber 
\\
&& \times 
\left( 178.9\, |G_1|^2 + 51.8\, {\rm Re}(G_1^*G_2) + 4.37\, |G_2|^2 \right),
\\
&& \Gamma[B^+ \to \bar D^0 D^{*0} K^+]  = 10^{-3} {\rm MeV} 
\nonumber 
\\
&& \times 
\left( 49.6\, |G_1|^2 + 2.61\, {\rm Re}(G_1^*G_2) + 3.49\, |G_2|^2 \right),
\\
&& \Gamma[B^+ \to \bar D^{*0} D^0 K^+]  = 10^{-3} {\rm MeV} 
\nonumber 
\\
&& \times 
\left( 52.5\, |G_1|^2 + 1.87\, {\rm Re}(G_1^*G_2) + 2.31\, |G_2|^2 \right),
\\
&& \Gamma[B^+ \to \bar D^{*0} D^{*0} K^+]  = 10^{-3} {\rm MeV} 
\nonumber 
\\
&& \times 
\left( 221.5\, |G_1|^2 + 74.8\, {\rm Re}(G_1^*G_2) + 11.58\, |G_2|^2 \right).
\end{eqnarray}
We obtain a good fit to the Babar branching fractions
with $G_1 = 3.2\times 10^{-6}$ and $G_2 = (-14.6 + 9.6i) \times 10^{-6}$.
In the corner of phase space where the 4-velocities of $\bar D^0$ 
and $D^{*0}$ are equal, the amplitudes
(\ref{A-Dbar*D}) and (\ref{A-DbarD*}) reduce to the form 
on the right side of (\ref{eq:A-DD*K}) with coefficients
$c_1 = c_2 = -i G_1/m_B + i G_2 (m_B+m_D+m_{D^*})/m_B^2$.
If $X$ has charge conjugation 
$C = +$, the estimate (\ref{eq:B-XK}) reduces to
\begin{eqnarray}
{\cal B}[B^+ \rightarrow X K^+] \approx 
\left( 2.6 \times 10^{-5} \right) {\Lambda_1^2 \over m_\pi^2}
\left( {E_b \over 0.4 \, {\rm MeV}} \right)^{1/2}.
\label{B-DDK:est}
\end{eqnarray}
If $C = -$, the branching fraction would be significantly smaller 
because of destructive interference between $c_1$ and $c_2$.  
We could get a more reliable result for the numerical prefactor 
in (\ref{B-DDK:est}) by relaxing the factorization assumption 
and carrying out a Dalitz plot analysis of the 3-body decays.
Since the result depends quadratically on the ultraviolet cutoff
$\Lambda_1$, the best we can do is obtain an order-of-magnitude
estimate of the branching fraction by setting 
$\Lambda_1 \approx m_\pi$. 

The Belle collaboration measured the product of the branching fractions
${\cal B}[B^+ \to X K^+]$ and ${\cal B}[X \to J/\psi \; \pi^+ \pi^-]$
to be $(1.3 \pm 0.3)\times 10^{-5}$ \cite{Choi:2003ue}.
Our estimate of ${\cal B}[B^+ \rightarrow X K^+]$ 
is compatible with this result
if $J/\psi \; \pi^+ \pi^-$ is one of the major decay modes of $X$.
The experimental upper bound on the width of $X(3872)$ is
$\Gamma_X < 2300$ keV.
The sum of the widths for decay into $\bar D^0 D^0 \pi^0$ 
and $\bar D^0 D^0 \gamma$ approaches $\Gamma[D^{*0}] \approx 50$ 
keV in the limit $E_b \to 0$ \cite{Voloshin:2003nt}.
The remaining partial widths scale as $E_b^{1/2}$.
Using a coupled-channel calculation in a model in which 
$X$ mixes with $J/\psi \;\rho$, the decay rate for $J/\psi \; \pi^+ \pi^-$
has been estimated to be 1290 keV for $E_b = 0.7$ MeV \cite{Swanson:2003tb}.
Thus it is at least plausible that $J/\psi \; \pi^+ \pi^-$ 
is one of the major decay modes.
Other possible decay channels are $\eta_c \pi \pi$,
radiative transitions to charmonium states,
and $c \bar c$ annihilation decays.

We have calculated the decay rate for
$B^+ \to X K^+$ and the differential
decay rate for $B^+ \to \bar{D}^0 D^{*0}K^+$
near the $\bar{D}^0 D^{*0}$ threshold
under the assumption that $X(3872)$ is a loosely-bound S-wave
$\bar{D}^0 D^{*0}$/$D^{0} \bar{D}^{*0}$ molecule
and that its production rate is dominated by the 
coalescense of charm mesons.
Observation of a sharp peak in the $\bar D^0 D^{*0}$ 
invariant mass distribution near threshold in the decay 
$B^+ \to \bar{D}^0 D^{*0}K^+$ would confirm the 
interpretation of $X$ as a $\bar D^0 D^{*0}$ molecule.
Our order-of-magnitude estimate of the branching fraction 
for $B^+ \to X K^+$ is compatible with observations 
if $X(3872)$ has quantum numbers $J^{PC} = 1^{++}$ and if 
$J/\psi \; \pi^+ \pi^-$ is one of its major decay modes.

This research was supported in part by the Department of Energy under
grant DE-FG02-91-ER4069.


\end{document}